\documentclass[12pt,letter]{article}
\usepackage{jcappub}

\def\gg{\hbox{\raise-2mm\hbox{$\textstyle \overline{g} \atop \scriptstyle 0$}}}
\def\g1{\hbox{\raise-2mm\hbox{$\textstyle \overline{g} \atop \scriptstyle 1$}}}
\def\uu{\hbox{\raise-2mm\hbox{$\textstyle u \atop \scriptstyle 0$}}}
\def\u1{\hbox{\raise-2mm\hbox{$\textstyle u \atop \scriptstyle 1$}}}
\def\ZZ{\hbox{\raise-2mm\hbox{$\textstyle Z \atop \scriptstyle 0$}}}
\def\Z1{\hbox{\raise-2mm\hbox{$\textstyle Z \atop \scriptstyle 1$}}}
\def\AA{\hbox{\raise-2mm\hbox{$\textstyle A \atop \scriptstyle 0$}}}
\def\A1{\hbox{\raise-2mm\hbox{$\textstyle A \atop \scriptstyle 1$}}}

\newcommand{\uline}[1]{\ensuremath{\underline{#1}}}
\newcommand{\ph}[1]{\ensuremath{\phantom{#1}}}
\newcommand{\al}{\alpha}
\newcommand{\ga}{\gamma}
\newcommand{\de}{\delta}
\newcommand{\ep}{\epsilon}
\newcommand{\ta}{\tau}
\newcommand{\si}{\sigma}
\newcommand{\rh}{\rho}
\newcommand{\la}{\lambda}

\newcommand{\be}{\begin{equation}}
\newcommand{\ee}{\end{equation}}
\newcommand{\bea}{\begin{eqnarray}}
\newcommand{\eea}{\end{eqnarray}}

\title{Observational Cosmology in Macroscopic Gravity}

\author[a]{Timothy\hspace{1mm}Clifton\hspace{1mm},\hspace{-3mm}}

\author[b]{\hspace{1mm}Alan\hspace{1mm}Coley}

\author[c]{\& Robert\hspace{1mm}van\hspace{1mm}den\hspace{1mm}Hoogen}

\affiliation[a]{School of Physics and Astronomy, Queen Mary University of London, UK.}
\affiliation[b]{Department of Mathematics and Statistics, Dalhousie University, Canada.}
\affiliation[c]{Department of Mathematics, Statistics and Computer Science,\\
  St. Francis Xavier University, Canada.}

\emailAdd{t.clifton@qmul.ac.uk}
\emailAdd{aac@mathstat.dal.ca}
\emailAdd{rvandenh@stfx.ca}

\abstract{
We discuss the construction of cosmological models within the framework of Macroscopic Gravity (MG), which is a theory that models
the effects of averaging the geometry of space-time on  large scales.  We find new exact spatially homogeneous and isotropic FLRW solutions to the MG field equations, and investigate large-scale perturbations around them. We find that any inhomogeneous perturbations to the averaged geometry are severely restricted, but that possible anisotropies in the correlation tensor can have dramatic consequences for the measurement of distances. These calculations are a first step within the MG approach toward developing averaged cosmological models to a point where they can be used to interpret real cosmological data, and hence to provide a working alternative to the ``concordance'' $\Lambda$CDM model.}

\arxivnumber{}

\begin{document}

\maketitle

\newpage

\section*{Introduction}

The currently favoured ``concordance model'' of cosmology treats the
entire observable universe as a single
Friedmann-Lema\^{i}tre-Robertson-Walker (FLRW) solution of
Einstein's equations.  The complex hierarchy of structures that astronomers
observe is then accommodated by allowing for small fluctuations to this
geometry, and a linearized version of Einstein's equations is employed
to determine their behaviour.  The mathematical simplicity of this
approach allows extremely complicated distributions of matter to be
considered, and permits one to perform detailed calculations of a
variety of complicated physical phenomena.  It relies, however, on
the assumption that the equations that govern the large-scale `average'
geometry of the Universe are the same as Einstein's equations.  This
will not be true in general.

What we have in reality is a complicated fitting problem:  We make
observations over various different scales (depending on the
particular observable in question), and we then look for the FLRW
model that best fits either one or all of these observables.  Ideally
it would be better to make direct observations of the matter
distribution, and then solve Einstein's equations with some reasonable
set of boundary conditions.  Unfortunately this is not currently possible, due to
the complexity of the equations, and so we are forced to revert to the
former approach.  In order for this procedure to have any meaning,
however, we must be prepared to answer a series of questions:  What is
the relation (if any) between the fitted FLRW model and the actual
geometry of space-time?  What are the evolution equations that the
fitted FLRW model should obey?  Should the FLRW models we fit to
observations made on different scales be expected to be the same, and, if
not, how (if at all) are they related?

One way to go about trying to answer these questions is to attempt to develop
and understand procedures for averaging the geometry of space-time,
thereby providing an explicit link between the microscopic geometry of
space-time, and the macroscopic averages that are used in cosmology.
This is the goal of Zalaletdinov's
theory of Macroscopic Gravity (MG)~\cite{Zalaletdinov},
which consists of a prescription for averaging tensors, vector and
scalars in a covariant way, as well as providing a set of field
equations that the averaged geometric quantities are expected to obey.
Despite the difficulties in MG \cite{vandenHoogen2008,vandenHoogen2009,A,C,D}, it does serve as a framework
within which calculations can be performed.
Highly symmetric exact solutions to the field equations of MG have
been found in~\cite{ColeyPelavasZalaletdinvov2005,ColeyPelavas,vandenHoogen2008,vandenHoogen2009},
the prescribed averaging procedure has been directly applied to
space-times that are already close to being spatially homogeneous and
istropic in~\cite{ps,ps1,ps2}, and a recent attempt to use the exact solutions found
in~\cite{vandenHoogen2009} and \cite{ColeyPelavasZalaletdinvov2005} to
interpret cosmological observations has been made in~\cite{CCCS}. For other approaches to averaging in cosmology the reader is referred to~\cite{A,B}, and references therein. The relationship between MG and some of these approaches is considered in~\cite{ps}.

Our aim here is to further develop the construction of cosmological models within the theory
of MG, thereby building on previous work that has identified exact spatially homogeneous and
isotropic solutions to the MG field equations.  In particular, we aim to investigate
cosmological perturbation theory within MG theory, as well as making some first steps
towards understanding the propagation of rays of light.  In Section \ref{MGtheory} we
briefly summarize the essentials of MG theory, and then proceed to
discuss its FLRW solutions in Section \ref{MGFLRW}.  We first recap some known solutions, before presenting new
solutions that have not previously appeared in the literature (details of these new
solutions can be found in the appendix).  In Section \ref{MGpert} we then investigate perturbations
around the known FLRW solutions, finding that MG theory requires that either: (i) The background
macroscopic geometry must be the same as the FLRW solutions of Einstein's equations (up to
the order of perturbations considered), or (ii) The perturbations themselves must be
spatially homogeneous (or take a very restricted form).  We then proceed in Section
\ref{Obs} to calculate measures of distance in the averaged geometry by using the average of the Ricci
and Weyl curvature tensors in the Sachs equations.  We find that in the simplest solutions
the correlation tensor acts as a spatial curvature term in the Sachs equations, in the same
way that it does in the MG field equations.  However, this result does not extend to more
general FLRW solutions which exhibit considerable freedom.  Finally, we conclude with some brief
remarks.

\section{Macroscopic Gravity Theory}
\label{MGtheory}

Macroscopic Gravity (MG) is Zalaletdinov's non-perturbative
approach to describing the behaviour of space-time on large
scales~\cite{Zalaletdinov}.  We will briefly
summarize it here.

\subsection{Covariant Averaging}

Zalaletdinov's approach involves averaging the geometric objects that exist on the space-time manifold, and constructing field
equations for these averaged quantities based on averaging the Cartan equations for the microscopic geometry. The first step in this is to
provide an averaging procedure that is covariant, and that maintains the tensorial
properties of the object that is being averaged.  To do this,
Zalaletdinov defines that the average of an object
$p^{\alpha \dots}_{\phantom{\alpha} \beta \dots}$, over some closed region of
space-time $\Sigma$ that contains the supporting point $x$, to be
\cite{Zalaletdinov}
\be
\left< p^{\alpha \dots}_{\beta \dots}(x) \right> =
\frac{1}{V_{\Sigma}} \int_{\Sigma} \sqrt{-g^{\prime}} d^4 x^{\prime}
p^{\mu^{\prime}\dots}_{\nu^{\prime} \dots} (x^{\prime})
\mathcal{A}^{\alpha}_{\phantom{\alpha} \mu^{\prime}} (x,x^{\prime})
 \mathcal{A}^{\nu^{\prime}}_{\phantom{\nu^{\prime}} \beta}
 (x,x^{\prime}) \dots,
\ee
where $V_{\Sigma}=\int_{\Sigma} \sqrt{-g} d^4 x$ is the 4-volume of
$\Sigma$, and $\mathcal{A}^{\alpha}_{\phantom{\alpha} \mu^{\prime}}
(x,x^{\prime})$ and $\mathcal{A}^{\nu^{\prime}}_{\phantom{\nu^{\prime}} \beta}
(x,x^{\prime})$ are bilocal averaging operators. These operators can be written as
the product of vector bases at two different points in $\Sigma$ as
$\mathcal{A}^{\alpha^{\prime}}_{\phantom{\alpha^{\prime}} \beta} (x,
x^{\prime}) = e^{\alpha^{\prime}}_{\phantom{\alpha^{\prime}} i} (x^{\prime})
e^i_{\phantom{i} \beta} (x)$, where the structure functions, $C_{i j}{}{^k}$ of  $[e_i,e_j]=C_{i j}{}^k e_k$ (i.e. the coefficients of anholonomity), are constants.


With this definition we can now consider the average of
various geometric objects.  A key point to bear in mind here is that
averaging non-linear quantities is not, in general, the same
thing as constructing those same quantities from the average of their arguments.
One must therefore proceed with care.  Let us
specify that we will write the un-averaged metric, connection and
Riemann tensor in the normal way as $g_{\mu \nu}$,
$\Gamma^{\mu}_{\phantom{\mu} \nu \rho}$ and $R^{\mu}_{\phantom{\mu}
  \nu \rho \sigma}$.   We can then write the averages of these quantities as
$\langle g_{\mu \nu} \rangle$, $\langle \Gamma^{\mu}_{\phantom{\mu}
  \nu \rho} \rangle$ and $\langle R^{\mu}_{\phantom{\mu} \nu \rho \sigma} \rangle$.
It is now, for example, no longer the case that the curvature tensor constructed
from $\langle \Gamma^{\mu}_{\phantom{\mu} \nu \rho} \rangle$ is given
by $\langle R^{\mu}_{\phantom{\mu} \nu \rho \sigma} \rangle$.  Instead,
following Zalaletdinov, we denote this object as
\be
M^{\mu}_{\phantom{\mu} \nu \alpha \beta} = \partial_{\alpha}
\langle\Gamma^{\mu}_{\phantom{\mu} \nu \beta}\rangle- \partial_{\beta}
\langle\Gamma^{\mu}_{\phantom{\mu} \nu \alpha}\rangle + \langle\Gamma^{\mu}_{\phantom{\mu}
  \sigma \alpha}\rangle \langle\Gamma^{\sigma}_{\phantom{\sigma} \nu \beta}\rangle -
\langle\Gamma^{\mu}_{\phantom{\mu}
  \sigma \beta}\rangle \langle\Gamma^{\sigma}_{\phantom{\sigma} \nu
  \alpha}\rangle,
\ee
where $M^{\mu}_{\phantom{\mu} \nu \alpha \beta} \neq \langle
R^{\mu}_{\phantom{\mu} \nu \alpha \beta} \rangle$, in general.
Likewise, it is assumed that there exists a non-metric compatible and symmetric connection, $\Pi^{\mu}_{\phantom{\mu} \nu
  \rho} $ \cite{Zalaletdinov}, such that
\be
\langle R^{\mu}_{\phantom{\mu} \nu \alpha \beta} \rangle = \partial_{\alpha}
\Pi^{\mu}_{\phantom{\mu} \nu \beta} - \partial_{\beta}
\Pi^{\mu}_{\phantom{\mu} \nu \alpha} + \Pi^{\mu}_{\phantom{\mu}
  \sigma \alpha} \Pi^{\sigma}_{\phantom{\sigma} \nu \beta} -
\Pi^{\mu}_{\phantom{\mu} \sigma \beta} \Pi^{\sigma}_{\phantom{\sigma} \nu
  \alpha},
\ee
where $\Pi^{\mu}_{\phantom{\mu} \nu \rho} \neq \langle
\Gamma^{\mu}_{\phantom{\mu} \nu \rho} \rangle$, in general.
The difference between these curvature tensors,
\begin{eqnarray}
\label{Qdef}
Q^{\mu}_{\phantom{\mu} \nu \alpha \beta}
&=& \langle R^{\mu}_{\phantom{\mu} \nu \alpha \beta} \rangle - M^{\mu}_{\phantom{\mu} \nu \alpha \beta}\\
&=& 2\langle\Gamma^{\mu}_{\phantom{\mu} \epsilon [\alpha}\Gamma^{\epsilon}_{\phantom{\epsilon} \underline{\nu}\beta]}\rangle - 2\langle\Gamma^{\mu}_{\phantom{\mu} \epsilon [\alpha}\rangle \langle\Gamma^{\epsilon}_{\phantom{\epsilon} \underline{\nu}\beta]}\rangle,
\end{eqnarray}
where under-lined indices are not included in anti-symmetrization, is known as the ``polarization tensor''. The difference between the averaged connection and the connection of the averaged Riemann tensor yields an ``affine deformation tensor'':
\begin{equation}
A^{\mu}_{\phantom{\mu} \nu \rho} = \langle\Gamma^{\mu}_{\phantom{\mu} \nu \rho} \rangle - \Pi^{\mu}_{\phantom{\mu} \nu \rho}.
\end{equation}
In what follows we will often use an over-bar on a quantity to denote the average, rather than angular brackets, as this will allow us to be more concise.

\subsection{The Macroscopic Field Equations}

Under the assumption that the connection
$\bar{\Gamma}^{\mu}_{\phantom{\mu} \nu \rho}=\langle
\Gamma^{\mu}_{\phantom{\mu} \nu \rho} \rangle$ is compatible with the
metric $\bar{g}_{\mu \nu}=\langle g_{\mu \nu} \rangle$, and that we assume the following splitting rules for products of connection and metric,
$\langle \Gamma^{\mu}_{\phantom{\mu} \nu \rho} g^{\sigma}_{\phantom{\sigma}
  \tau}  \rangle =  \langle \Gamma^{\mu}_{\phantom{\mu} \nu \rho}
\rangle \langle g^{\sigma}_{\phantom{\sigma} \tau} \rangle$ and
$\langle
\Gamma^{\alpha}_{\phantom{\alpha} \beta [ \gamma}
\Gamma^{\mu}_{\phantom{\mu} \underline{\nu} \rho]} g^{\sigma}_{\phantom{\sigma}
  \tau}  \rangle =  \langle \Gamma^{\alpha}_{\phantom{\alpha} \beta [ \gamma}
\Gamma^{\mu}_{\phantom{\mu} \underline{\nu} \rho]}\rangle
\langle g^{\sigma}_{\phantom{\sigma} \tau} \rangle$, one can
then construct the field equations of MG:
\be
\label{MFE}
\bar{g}^{\beta \epsilon } M_{\gamma \beta} - \frac{1}{2}
\delta^{\epsilon}_{\phantom{\epsilon} \gamma} \bar{g}^{\mu \nu}
M_{\mu \nu} = 8 \pi G \bar{T}^{\epsilon}_{\phantom{\epsilon}
  \gamma} - (Z^{\epsilon}_{\phantom{\epsilon} \mu \nu
  \gamma} - \frac{1}{2} \delta^{\epsilon}_{\phantom{\epsilon} \gamma}
Q_{\mu \nu} ) \bar{g}^{\mu \nu},
\ee
where $\bar{T}^{\epsilon}_{\phantom{\epsilon} \gamma}$ is the
averaged energy-momentum tensor, and where $Z^{\alpha}_{\phantom{\alpha}
  \mu \nu \beta}=2 Z^{\alpha \phantom{\mu \epsilon}
    \epsilon}_{\phantom{\alpha} \mu \epsilon \phantom{\epsilon} \nu
  \beta }$ and $Q_{\mu \nu}= Z^{\alpha}_{\phantom{\alpha} \mu \nu \alpha} $. Here $Z^{\alpha \phantom{\beta  \gamma} \mu}_{\phantom{\alpha} \beta
      \gamma \phantom{\mu} \nu \sigma }$ is known as the
  ``connection correlation tensor'', and is given by
\be
\label{Zdef}
Z^{\alpha \phantom{\beta  \gamma} \mu}_{\phantom{\alpha} \beta
      \gamma \phantom{\mu} \nu \sigma }
=
\langle \Gamma^{\alpha}_{\phantom{\alpha} \beta [ \gamma}
  \Gamma^{\mu}_{\phantom{\mu} \underline{\nu} \sigma ]} \rangle -
\langle \Gamma^{\alpha}_{\phantom{\alpha} \beta [ \gamma} \rangle
  \langle \Gamma^{\mu}_{\phantom{\mu} \underline{\nu} \sigma ]} \rangle,
\ee
 We note that the ``polarization tensor'' is the trace of the ``connection correlation tensor'', via $Q^\alpha_{\phantom{\alpha}\beta\mu\nu}=-2Z^{\epsilon\phantom{\beta\mu}\alpha}_{\phantom{\epsilon}\beta\mu\phantom{\alpha}\epsilon\nu}$. We assume that $\bar{T}^{\mu}_{\phantom{\mu} \nu}=\langle T^{\mu}_{\phantom{\mu} \nu} \rangle$ can be written as a perfect fluid, such that
\be
\label{MGT}
\bar{T}^{\mu}_{\phantom{\mu} \nu} = (\rho+p) u^{\mu} u_{\nu}
+ p \bar{g}^{\mu}_{\phantom{\mu} \nu} +
\Sigma^{\mu}_{\phantom{\mu}\nu},
\ee
where $\rho$, $p$ and $\Sigma^{\mu}_{\phantom{\mu} \nu}$ are the energy density, isotropic pressure, and anisotropic stress of the fluid, respectively.  The 4-vector $u^{\mu}$ exists in the manifold with averaged geometry $\bar{g}_{\mu \nu}$, and is the velocity 4-vector of the fluid.

\subsection{The Connection Correlation Tensor}

From the definition of the connection correlation tensor (\ref{Zdef}), it can be seen that
$Z^{\alpha \phantom{\beta  \gamma} \mu}_{\phantom{\alpha} \beta (\gamma \phantom{\mu} \underline{\nu} \sigma )}=0 $,
$Z^{\alpha \phantom{\beta\gamma} \mu}_{\phantom{\alpha} \beta \gamma \phantom{\mu} \nu\sigma } =
-Z^{\mu \phantom{\nu  \gamma} \alpha}_{\phantom{\mu} \nu \gamma \phantom{\alpha} \beta \sigma }$, and that we have the cyclic constraint
\be
\label{cyclic}
Z^{\alpha\phantom{\beta  \gamma} \mu}_{\phantom{\alpha} \beta [ \gamma\phantom{\mu} \nu \sigma ]}=0.
\ee
The connection correlation tensor
is also assumed to satisfy the ``equi-affinity'' constraint
\be
\label{equi-Z}
Z^{\epsilon \phantom{\epsilon  \gamma} \mu}_{\phantom{\epsilon} \epsilon\gamma \phantom{\mu} \nu \sigma }=0
\ee
and the
differential constraint
\be
\label{Zdif}
Z^{\alpha
  \phantom{\beta [ \gamma} \mu}_{\phantom{\alpha} \beta[ \gamma
      \phantom{\mu} \underline{\nu} \sigma \vert \vert \lambda]}=0,
\ee
where $\vert \vert$ denotes a covariant derivative with respect to the averaged
connection, $\bar{\Gamma}^{\mu}_{\phantom{\mu} \nu \rho}$.
The integrability conditions for this equation are
\be
  Z{}^\epsilon{}_{\beta    [\mu }{}^\gamma   {}_{{\underline\delta}
      \nu} M{}^\alpha   {}_{{\underline\epsilon}  \kappa\pi]}
- Z{}^\alpha  {}_{\epsilon [\mu }{}^\gamma   {}_{{\underline\delta}
    \nu} M{}^\epsilon {}_{{\underline\beta}     \kappa\pi]}
+ Z{}^\alpha  {}_{\beta    [\mu }{}^\epsilon {}_{{\underline\delta}
    \nu} M{}^\gamma   {}_{{\underline\epsilon}  \kappa\pi]}
- Z{}^\alpha  {}_{\beta    [\mu }{}^\gamma   {}_{{\underline\epsilon}
    \nu} M{}^\epsilon {}_{{\underline\delta}    \kappa\pi]}=0.
\label{ZM}
\ee
With the assumption of vanishing 3 and 4-form correlation functions for the
connection and assuming Eq. (\ref{Zdif}),
we have the quadratic constraint
\begin{eqnarray}
&&Z^{\delta }{}_{\beta \lbrack \gamma }{}^{\theta
}{}_{\underline{\kappa }\pi }Z^{\alpha }{}_{\underline{\delta
}\epsilon }{}^{\mu }{}_{\underline{\nu } \sigma ]}+
Z^{\delta }{}_{\beta \lbrack \gamma }{}^{\mu }{}_{\underline{\nu } \sigma
}Z^{\theta }{}_{\underline{\kappa }\pi }{}^{\alpha }{}_{\underline{
\delta }\epsilon ]}
\nonumber \\
+&&Z^{\alpha }{}_{\beta \lbrack \gamma }{}^{\delta }{}_{ \underline{\nu
}\sigma }Z^{\mu }{}_{\underline{\delta }\epsilon }{}^{\theta
}{}_{\underline{\kappa }\pi ]}+
Z^{\alpha }{}_{\beta \lbrack \gamma }{}^{\mu
}{}_{\underline{\delta } \epsilon }Z^{\theta }{}_{\underline{\kappa
}\pi }{}^{\delta }{}_{\underline{ \nu }\sigma ]}
\nonumber \\
+&&Z^{\alpha}{}_{\beta \lbrack \gamma }{}^{\theta }{}_{ \underline{\delta
}\epsilon }Z^{\mu }{}_{\underline{\nu }\sigma }{}^{\delta
}{}_{\underline{\kappa }\pi ]}+
Z^{\alpha }{}_{\beta \lbrack \gamma }{}^{\delta
}{}_{\underline{\kappa }\pi }Z^{\theta }{}_{\underline{\delta }
\epsilon }{}^{\mu }{}_{\underline{\nu }\sigma ]}=0.
\label{ZZ}
\end{eqnarray}
Eqs. (\ref{Zdif})-(\ref{ZZ})
are then solved to determine the connection correlation tensor.

\subsection{The Affine Deformation Tensor}

The affine deformation tensor must obey the differential constraint
\be
\label{Adif}
A^{\alpha}_{\phantom{\alpha} \beta [ \nu \vert \vert \mu]} -
A^{\alpha}_{\phantom{\alpha} \epsilon [ \mu}
  A^{\epsilon}_{\phantom{\epsilon} \underline{\beta} \nu]}
= -\frac{1}{2} Q^{\alpha}_{\phantom{\alpha} \beta \mu \nu}.
\ee
From the Bianchi identities, and Eqs. (\ref{Qdef}) and (\ref{Zdif}), we then have
$\bar{R}^{\alpha}_{\phantom{\alpha} \beta [ \rho \sigma \vert \vert \lambda
]}=0$, where $\bar{R}^{\alpha}_{\phantom{\alpha} \beta \rho \sigma} =
\langle R^{\alpha}_{\phantom{\alpha} \beta \rho \sigma} \rangle$, which gives
\be
\label{AR}
A^{\epsilon}_{\phantom{\epsilon} \beta [ \rho}
  \bar{R}^{\alpha}_{\phantom{\alpha} \underline{\epsilon} \sigma \lambda ]}
-
A^{\alpha}_{\phantom{\alpha} \epsilon [ \rho}
  \bar{R}^{\epsilon}_{\phantom{\epsilon} \underline{\beta} \sigma \lambda
]}=0.
\ee
Eqs. (\ref{Adif})-(\ref{AR}) are solved to determine the affine
deformation tensor.

\section{Macroscopic FLRW Solutions}
\label{MGFLRW}

The FLRW solutions to Eqs. (\ref{MFE})-(\ref{AR}) have been
studied by Coley, Pelavas and
Zalaletdinov~\cite{ColeyPelavasZalaletdinvov2005}, and van den
Hoogen~\cite{vandenHoogen2009}.  The latter of these gave a systematic
and explicit investigation of all spatially flat solutions
in which $\bar{g}_{\mu \nu}$, $Z^{\alpha \phantom{\beta \gamma}
  \mu}_{\phantom{\alpha} \beta \gamma \phantom{\mu} \nu \sigma }$, and
$A^{\mu}_{\phantom{\mu} \nu \sigma}$ are invariant under the full
six-dimensional group of Killing vectors that characterize the
spatially homogeneous and isotropic FLRW geometries.  We
will now briefly outline these solutions (which we will call Type I), as
well as some more general FLRW solutions in which the connection correlation and
affine deformation tensors no longer have the same symmetries as $\bar{g}_{\mu\nu}$ (which we will call Type II).  These
latter solutions have not appeared in the literature before now.

\subsection{Type I Solutions}
\label{TypeI}

These solutions have correlation and affine deformation tensors that
are invariant under the same group of Killing vectors as
$\bar{g}_{\mu\nu}$.  In their most general form they are fully
specified by three independent constants: $\mathcal{A}$, $h_2$ and
$b_1$.  All three of these parameters appear in the (constant)
components of the connection correlation tensor, with the affine deformation
tensor being fully specified by $\mathcal{A}$ only (see \cite{vandenHoogen2009} for
details).

Writing the macroscopic line-element as
\be
d\bar{s}^2 = \bar{g}_{\mu\nu}dx^{\mu} dx^{\nu} = -dt^2
  + a^2(t) \left[dx^2+dy^2+dz^2 \right],
\ee
we then have that the macroscopic field equations (\ref{MFE}) give
\be
\label{MF1}
\frac{\dot{a}^2}{a^2} = \frac{8 \pi G}{3} \rho -
\frac{\mathcal{A}^2}{a^2},
\ee
where over-dots denote differentiation with respect to $t$, and where
$\rho$ is the macroscopic energy density from Eq. (\ref{MGT}) that obeys
\be
\label{MF2}
\dot{\rho} + 3 \frac{\dot{a}}{a} (\rho+p)=0.
\ee
The macroscopic geometry therefore evolves like the FLRW solutions of
Einstein's equations with an extra spatial curvature term in the
macroscopic equivalent of the Friedmann equation, even though the macroscopic geometry is spatially flat.

The constants $h_2$ and $b_1$ do not affect the dynamics of the averaged
space-time, and with $h_2=b_1=0$ the non-trivial components of the connection
correlation and affine deformation tensors are
\begin{eqnarray}
&&Z{}^j{}_{jj}{}^k{}_{jk} = Z{}^j{}_{kj}{}^k{}_{kk} =
  Z{}^j{}_{jj}{}^j{}_{kk} = Z{}^k{}_{jj}{}^k{}_{kk} = -\frac{2}{10} \mathcal{A}^2\\
&&Z{}^i{}_{jj}{}^k{}_{ik} = Z{}^j{}_{ij}{}^i{}_{kk} =
  Z{}^i{}_{jj}{}^i{}_{kk} = Z{}^j{}_{ij}{}^k{}_{ik} = -\frac{1}{10} \mathcal{A}^2,
\label{soln}
\end{eqnarray}
where $[ijk]$ is chosen to be one of the ordered triples
$\{[1,2,3],[2,3,1],[3,1,2]\}$, and
\be
A{}^0{}_{ii}= \mathcal{A}a(t),\qquad  A{}^i{}_{i0}= A{}^0{}_{00} = \frac{\mathcal{A}}{a(t)}, 
\ee
where $i$ takes on values $1,2,3$.  The observational consequences of a simple application of this solution have been investigated by Clarkson {\it et al.} \cite{CCCS}.

\subsection{Type II Solutions}
\label{TypeII}

Further solutions with a FLRW macroscopic metric,
$\bar{g}_{\mu \nu}$, can be found if $Z^{\alpha \phantom{\beta \gamma}
  \mu}_{\phantom{\alpha} \beta \gamma \phantom{\mu} \nu \sigma }$ and
$A^{\mu}_{\phantom{\mu} \nu \sigma}$ are allowed to be less symmetric.  From
Eq. (\ref{MFE}) it can be seen that this is possible because only
certain contractions of these tensors are required to admit the full
six-dimensional group of Killing vectors that are required for $\bar{g}_{\mu
  \nu}$ to be an FLRW metric.  That is, the correlation and affine deformation
tensors do {\it not} have to be exhibit invariance under
all spatial translations and rotations themselves in order for $\bar{g}_{\mu \nu}$ to be FLRW.

We find that, by relaxing the symmetry requirements on $Z^{\alpha
  \phantom{\beta \gamma}
  \mu}_{\phantom{\alpha} \beta \gamma \phantom{\mu} \nu \sigma }$ and
$A^{\mu}_{\phantom{\mu} \nu \sigma}$, there exists a family of four solutions to
the algebraic constraint Eqs. (\ref{ZM}), (\ref{ZZ}) and (\ref{AR}) that are
parameterized by the ordered pair $(\alpha_1,\alpha_2)\in \{(1/2,0),(1/2,1/2),(1,0), (1,1/2)\}$.
In this case it is found that $Z^{\alpha \phantom{\beta \gamma} \mu}_{\phantom{\alpha}
  \beta \gamma \phantom{\mu} \nu \sigma }$ contains:
\be
{\rm 51} \;  {\rm functions} \; f_i(x,y,z), \; {\rm 2} \; {\rm
  functions} \; \mathcal{F}_i(x,y), \; {\rm and} \; {\rm 1} \; {\rm
  constant} \; \mathcal{A},
\ee
and $A^{\mu}_{\phantom{\mu} \nu \sigma}$ contains:
\be
{\rm 2} \;  {\rm functions} \; \mathcal{B}_i(x,y) \; {\rm and} \; {\rm 2} \; {\rm constants} \; \mathcal{A}, \mathcal{B}.
\ee
where $\mathcal{A}$ is the same constant as found in the Type I solutions, in the appropriate limit.  Equation (\ref{Zdif}) then provides a lengthy set of differential equations that must be obeyed by the $f_i(x,y,z)$, and Eq. (\ref{Adif}) provides the differential constraints (\ref{A_dif2}),
which essentially define the connection correlation functions $\mathcal{F}_i(x,y)$ in terms of the functions $\mathcal{B}_i(x,y)$ found in the Affine Deformation tensor.

The macroscopic field equations (\ref{MFE}) then reduce to the same
form as in the Type I solutions described above, such that
Eqs. (\ref{MF1}) and (\ref{MF2}) must be satisfied.  That is, the only
quantity that enters into the macroscopic metric $\bar{g}_{\mu \nu}$
from the correlation tensor is the single constant $\mathcal{A}$,
which appears as a spatial curvature term (this can be seen to be due to assumption 4 in the appendix).  The extra freedom
introduced into the macroscopic description of space-time by allowing
$Z^{\alpha \phantom{\beta \gamma}
  \mu}_{\phantom{\alpha} \beta \gamma \phantom{\mu} \nu \sigma }$ and
$A^{\mu}_{\phantom{\mu} \nu \sigma}$ to be less symmetric than
$\bar{g}_{\mu\nu}$ does not therefore, in this case, result in any
different behaviour in the macroscopic metric.  It does, however,
result in extra terms in both the connection correlation tensor and affine
deformation tensor that we will find can be important for observations
made within the space-time.

The result that the macroscopic metric obeys
Eq. (\ref{MF1}) is true independent of whether or not we set the $f_i(x,y,z)$ to be constants.
For further details of this generalized solution, the reader is
referred to the appendix.

\section{Perturbed Spatially Homogeneous and Isotropic Solutions}
\label{MGpert}

We now want to look for solutions to the equations of macroscopic
gravity, as outlined in Section \ref{MGtheory}, that are less
symmetric than the spatially homogeneous and isotropic macroscopic
geometries that were studied in Section \ref{MGFLRW}.  We will do this
by allowing for small fluctuations about the FLRW geometries that we
already know, and then performing a perturbative expansion in the
fluctuations at the level of the equations that govern
$\bar{g}_{\mu\nu}$, $Z^{\alpha \phantom{\beta \gamma}
  \mu}_{\phantom{\alpha} \beta \gamma \phantom{\mu} \nu \sigma }$ and
$A^{\mu}_{\phantom{\mu} \nu \sigma}$.  This method is routinely used
when studying the `almost' FLRW solutions that are heavily relied
upon in the standard cosmological model.  It is within this type
of frame-work that predictions involving weak lensing, galaxy correlation
functions, and CMB observations are routinely calculated.
As the MG equations describe macroscopic behaviour, we shall only consider fluctuations on scales comparable to or larger than the averaging domains $\Sigma$. In what follows we will refer to these as `macroscopic scales'.

\subsection{Perturbed Macroscopic Variables}

Let us begin by writing the perturbed macroscopic geometry as
\be
\label{pertFLRW}
d\bar{s}^2 = a^2(\eta) \left[- \left( 1+2 \Phi \right) d\eta^2 + \left( 1
  - 2 \Psi \right) \left( dx^2 + dy^2 + dz^2 \right) \right],
\ee
where $\eta = \int dt/a(t)$ is the conformal time, and is
used here to simplify the equations.  The perturbed velocity 4-vector
can then be written as
\be
u^{\mu}= \frac{1}{a} \left( 1-\Phi, v^i \right),
\ee
where the appearance of $\Phi$ in the temporal component is due to the
normalization $u^{\mu}u_{\mu}=-1$, and $v^i$ is the peculiar velocity.  For the perturbed energy-momentum tensor we then write
\begin{eqnarray}
T^{0}_{\phantom{0} 0} &=& - \delta \rho\\
T^{0}_{\phantom{0} i} &=& - (\rho+p) v_i\\
T^{i}_{\phantom{i} j} &=& \delta^{i}_{\phantom{i} j} \delta p + (\rho+p) D^{i}_{\phantom{i} j} \Sigma,
\end{eqnarray}
where $D_{i j} = \partial_i \partial_j - \delta_{ij} \nabla^2$ is a traceless spatial derivative operator, $\delta \rho$ is the energy density perturbation, $\delta p$ is the pressure perturbation, and $\Sigma$ is the scalar anisotropic stress.
We further assume that we can write
\begin{eqnarray}
\label{Zpert}
Z^{\alpha \phantom{\beta \gamma}
  \mu}_{\phantom{\alpha} \beta \gamma \phantom{\mu} \nu \sigma } &=& {Z^{(0)}}^{\alpha \phantom{\beta \gamma}
  \mu}_{\phantom{\alpha} \beta \gamma \phantom{\mu} \nu \sigma }  + \delta Z^{\alpha \phantom{\beta \gamma}
  \mu}_{\phantom{\alpha} \beta \gamma \phantom{\mu} \nu \sigma }\\
A^{\mu}_{\phantom{\mu} \nu \sigma} &=& {A^{(0)}}^{\mu}_{\phantom{\mu}
  \nu \sigma} + \delta A^{\mu}_{\phantom{\mu} \nu \sigma},
\label{pertA}
\end{eqnarray}
where a superscript $(0)$ denotes the value of a quantity in the exact
FLRW solution that is being perturbed, and 
$\delta Z^{\alpha \phantom{\beta \gamma}
  \mu}_{\phantom{\alpha} \beta \gamma \phantom{\mu} \nu \sigma }$ and
$\delta A^{\mu}_{\phantom{\mu} \nu \sigma}$ indicate small
fluctuations. In general, $\delta Z^{\alpha \phantom{\beta \gamma}
  \mu}_{\phantom{\alpha} \beta \gamma \phantom{\mu} \nu \sigma }$
has 720 independent components, and $\delta A^{\mu}_{\phantom{\mu}
  \nu \sigma}$ has 40 independent components.

\subsection{Perturbed Equations}
\label{perteqns}

We now want to substitute our perturbed quantities (as specified in
Eqs. (\ref{pertFLRW})-(\ref{pertA})) into
Eqs. (\ref{Zdif})-(\ref{AR}), and to expand the result
in some `order of smallness' parameter, $\epsilon$.  In doing this, we take our
fluctuations to be of the following order:
\be
\Phi \sim \Psi \sim v^i \sim \delta \rho \sim \delta p \sim \Sigma \sim \delta
Z^{\alpha \phantom{\beta \gamma} \mu}_{\phantom{\alpha} \beta \gamma
  \phantom{\mu} \nu \sigma } \sim \delta A^{\mu}_{\phantom{\mu} \nu
  \sigma} \sim \epsilon.
\ee
The $O(0)$ part of these equations is automatically satisfied, if we
perturb around the exact FLRW solutions that we presented in Section
\ref{MGFLRW}.  We are then left with a set of equations that the
fluctuations must obey.

The general form of the perturbed equations is very lengthy, with a
large number of functions describing the fluctuations (in particular,
in $\delta Z^{\alpha \phantom{\beta \gamma} \mu}_{\phantom{\alpha}
  \beta \gamma \phantom{\mu} \nu \sigma }$ and $\delta
A^{\mu}_{\phantom{\mu} \nu \sigma}$).  The variables that we are most
interested in, however, are those that are involved in the macroscopic
geometry and the macroscopic fluid description of the matter, i.e. $\Phi$, $\Psi$, $\delta \rho$, $\delta p$, $\Sigma$ and $v^i$.
Information can be gained on the form that these fluctuations must take in the general
case, but in order to illustrate the form of the perturbed equations
in this paper we will first consider some simplifying special cases.

Let us begin by considering small fluctuations around the Type I
solutions from Section \ref{TypeI}. We can set some of the fluctuation in the correlation tensor to zero by choosing
\be
\label{dzE0}
\delta Z^{\alpha \phantom{\beta \gamma} \mu}_{\phantom{\alpha}
  \beta \gamma \phantom{\mu} \nu \sigma } u^{\sigma} =0.
\ee
In the language of \cite{vandenHoogen2009}, this corresponds to
setting the electric components of the correlation tensor to zero.
This condition means that the quadratic constraint, Eq. (\ref{ZZ}), is
automatically satisfied.  It also reduces the number of independent
components in $\delta Z^{\alpha \phantom{\beta \gamma} \mu}_{\phantom{\alpha}
  \beta \gamma \phantom{\mu} \nu \sigma }$ to 360.
The cyclic and equi-affine constraints, $Z^{\alpha
  \phantom{\beta  \gamma} \mu}_{\phantom{\alpha} \beta [ \gamma
    \phantom{\mu} \nu \sigma ]}=0$ and $Z^{\alpha \phantom{\alpha \gamma} \mu}_{\phantom{\alpha}
  \alpha \gamma \phantom{\mu} \nu \sigma }=0$, together with the
integrability conditions, Eq. (\ref{ZM}), and the differential constraint equation of the affine deformation tensor, Eq. (\ref{Adif}), then leave only 64
independent variables in $\delta
Z^{\alpha \phantom{\beta \gamma} \mu}_{\phantom{\alpha} \beta \gamma
  \phantom{\mu} \nu \sigma }$.

\subsubsection{The Macroscopic Field Equations}

After lengthy calculation, the macroscopic field equations (\ref{MFE}), at first order in perturbations, are found to give
\begin{eqnarray}
\label{MFE1}
\left(\frac{\partial^2} {(\partial {x^i})^2} -
\frac{\partial^2 }{(\partial {x^j})^2}\right) \left[ \Phi-\Psi
- 8 \pi G a^2 (\rho+p) \Sigma \right] &=& 0 \\
\label{MFE2}
\frac{\partial^2}{\partial x^i \partial x^j} \left[ \Phi-\Psi- 8 \pi G a^2 (\rho+p) \Sigma \right]&=& 0,
\end{eqnarray}
where $i \neq j$, and no sum is implied in Eq. (\ref{MFE1}). 
We also find
\begin{eqnarray}
\label{const1}
\nabla^2 \Phi -3 \mathcal{H} \left( \Phi^{\prime}+\mathcal{H} \Psi \right)&=& 4 \pi G a^2 \left(\delta \rho + \delta \rho_{\rm grav} \right)\\
\frac{\partial}{\partial x^i} \left( \Phi^{\prime}  + \mathcal{H} \Psi \right)
&=& 4 \pi G a^2 \left( \rho+ p -2 \frac{\mathcal{A}^2}{a^2}  \right) v_i
\label{const2}\\
\Phi^{\prime \prime} + \mathcal{H} \Psi^{\prime} +2 \mathcal{H} \Phi^{\prime} + \left( 2 \mathcal{H}^{\prime} +\mathcal{H}^2 \right) \Psi &+& \frac{1}{3} \nabla^2 (\Psi-\Phi) = 4 \pi G a^2 \left( \delta p -\frac{1}{3} \delta \rho_{\rm grav} \right), \qquad
\label{const3}
\end{eqnarray}
where a prime denotes differentiation with respect to $\eta$, and we have defined $\mathcal{H} = a^{\prime}/a$.  In Eqs. (\ref{const1})-(\ref{const3}) we have introduced the new quantity $\delta \rho_{\rm grav}$, which contains all the first-order contributions from the perturbed correlation tensor, with
 the exception of the term containing $\mathcal{A}$ in Eq. (\ref{const2}).

\subsubsection{The Connection Correlation Tensor}

Moving on from the macroscopic field equations, the integrability condition (\ref{ZM}) also means that $\Phi$ must obey the following equations:
\begin{eqnarray}
\label{pertMZ1a}
\mathcal{A}^2 \left[ \frac{\partial^2 \Phi}{(\partial {x^i})^2} -
\frac{\partial^2 \Phi}{(\partial {x^j})^2} \right] &=& 0  \qquad \qquad [{\rm no \; sum \; implied}]\\ \label{pertMZ1b}
\mathcal{A}^2 \frac{\partial^2 \Phi}{\partial x^i \partial x^j} &=& 0,
\end{eqnarray}
where $i\neq j$ and $i$, $j \in \{1,2,3\}$.  The differential constraint equations for the correlation tensor, (\ref{Zdif}), then give
\begin{eqnarray}
\label{pertMZ2a}
\mathcal{A}^2 \left[ \frac{\partial v_i}{\partial x^i} - \frac{\partial v_j}{\partial x^j} \right] &=& 0 \qquad \qquad [{\rm no \; sum \; implied}]\\
\mathcal{A}^2 \frac{\partial v_i}{\partial x^j} &=& 0,
\label{pertMZ2b}
\end{eqnarray}
where $i\neq j$.  Note that we have not yet assigned
an order of smallness to $\mathcal{A}$, so if $\mathcal{A} \sim
O(\epsilon^{1/2})$ or smaller then Eqs. (\ref{pertMZ1a})-(\ref{pertMZ2b})
are automatically satisfied.

Beyond these equations, we also gain from Eqs. (\ref{Zdif}) and (\ref{ZM}) constraint equations for $\Psi^{\prime}$ and $v_i^{\prime}$, as well 42 linearly independent equations involving spatial derivatives of the first-order contributions to the correlation tensor.

\subsubsection{The Affine Deformation Tensor}

At first order in perturbations there are initially 40 independent components to ${\delta A}^\alpha{}_{\beta\gamma}$. The linearized version of Eq. (\ref{AR}), assuming Eqs. (\ref{pertMZ1a}) and (\ref{pertMZ2b}), fixes all but 12.

\subsection{Macroscopic Perturbed FLRW Solutions}

Having summarized the equations for first-order perturbations about an FLRW background in MG, let us now turn to their solutions.

\subsubsection{The Macroscopic Field Equations}

The off-diagonal spatial components of the macroscopic field equations, given in Eqs. (\ref{MFE1}) and (\ref{MFE2}), can be seen to take exactly the same form as they do in the linearly perturbed FLRW solutions of Einstein's equations.  These equations are usually taken to imply that $\Phi-\Psi=8\pi G a^2 (\rho+p) \Sigma$, so that when the anisotropic stress vanishes we have $\Phi=\Psi$.  In the most general situation, we can see that $\Phi-\Psi- 8\pi G a^2 (\rho+p) \Sigma$ must be of the form
\begin{equation}
c_1 + c_2 x + c_3 y + c_4 z + c_5 (x^2+y^2+z^2),
\label{form}
\end{equation}
where $c_1$, $c_2$, $c_3$, $c_4$ and $c_5$ are all functions of $\eta$ only, and are of first order in our perturbative expansion.

Turning to Eqs. (\ref{const1})-(\ref{const3}) we again see that these equations take a similar form to the familiar ones that are arrived at when considering perturbed FLRW solutions of Einstein's equations.  The contributions to these equations from the correlation tensor occurs solely in the appearance of the terms containing $\delta \rho_{\rm grav}$ and $\mathcal{A}$.  This is a considerable increase in simplicity over the large number of variables that are required in order to specify the correlation tensor itself.  Moreover, it can be seen that the contribution of the correlation tensor to the macroscopic field equations takes the form of a perfect fluid with effective energy density and pressure of the form
\begin{eqnarray}
\rho_{\rm grav} &=& -\frac{3 \mathcal{A}^2}{a^2} + \delta \rho_{\rm grav} +O(\epsilon^2)\\
p_{\rm grav} &=& \frac{\mathcal{A}^2}{a^2} - \frac{1}{3} \delta \rho_{\rm grav} + O(\epsilon^2).
\end{eqnarray}
Clearly, this effective fluid has the following equation of state:
\be
p_{\rm grav} = -\frac{1}{3} \rho_{\rm grav} + O(\epsilon^2).
\ee
This is same effective equation of state as that of a spatial curvature term
in the Friedmann equations, and so the result that the correlation tensor mimicks spatial curvature is found here to be valid not only at the level of the background, but also to at least first order in perturbations around the background.  This extends the findings of \cite{ColeyPelavasZalaletdinvov2005} and \cite{vandenHoogen2009} beyond perfectly spatially homogeneous and isotropic solutions.

\subsubsection{The Connection Correlation Tensor}

The differential constraint (\ref{Zdif}), and its integrability condition (\ref{ZM}), provide us at first order in perturbations with Eqs. (\ref{pertMZ1a})-(\ref{pertMZ2b}).  If we write $v_i=v^{(s)}_i + v^{(v)}_i$,  where $v^{(s)}_i$ is the curl-free part of $v_i$ and $v^{(v)}_i$ is the divergence-free part of $v_i$, then we can also write $v^{(s)}_i = \partial \Theta / \partial x^i$.  If $\mathcal{A}\sim O(1)$ then Eqs. (\ref{pertMZ1a})-(\ref{pertMZ2b}) show that $\Theta$ and $\Phi$ must take the form given in Eq. (\ref{form}), and if $\Phi=\Psi$ then the same must also be true for $\Psi$.  These equations are therefore very constraining on the form that $\Phi$, $\Psi$ and the scalar part of $v_i$ can take.  In fact, if we were to perform the usual treatment of these quantities in terms of Fourier modes, then the resulting equations would suggest that all modes vanish to first order in perturbations.  That is, they would show that no inhomogeneous perturbations are possible at all.

\subsection{Interpretation}

First of all, one may consider what happens to the perturbations when the assumptions we
have imposed on the solution are relaxed.  For example, we could have used the more general
Type II solutions found in Section \ref{TypeII}.  Alternatively, we could have attempted to
relax our assumptions involving the vanishing of the electric part of $Z$, as prescribed by
Eq.  (\ref{dzE0}).  It is our contention that in these cases one still obtains equations
similar to (\ref{pertMZ1a})-(\ref{pertMZ2b}).  This is due to the form of Eq.  (\ref{ZM}),
which consists of terms built from contractions of the correlation tensor, $Z$, with the
Riemann tensor constructed from the averaged connection, $M$.  At first-order in
perturbations we will therefore generically have equations containing terms that contain
the zero-order part of $Z$ and the first-order part of $M$.  It is exactly these terms that
lead to Eqs.  (\ref{pertMZ1a}) and (\ref{pertMZ1b}), and which should therefore be
considered generic.  We note here that an exceptional case is the one in which the
background space-time is Minkwoski space, as in this case the zero-order part of $Z$
vanishes, and Eqs.  (\ref{pertMZ1a})-(\ref{pertMZ2b}) are satisfied identically.  This is
the reason that it was possible to find non-trivial spherically symmetric solutions to the MG equations in \cite{vandenHoogen2008} (see also  \cite{ColeyPelavas}).

It could perhaps be argued that Eqs.  (\ref{pertMZ1a})-(\ref{pertMZ2b}) only need to
be satisfied in some `averaged' sense, but this is incompatible with the approach taken here,
and would, presumably, require some adjustment of the macroscopic theory.  The equations we
have obtained consequently must be taken as constraints on the form of the macro metric
(\ref{pertFLRW}), and as such need to be interpreted.

If we assume $Z$ can be perturbed as
in Eq.  (\ref{Zpert}), and that the MG approach is generally valid, then we see 2 general possibilities:

\begin{itemize}

\item[1).] $\mathcal{A} \simeq 0$.  That is, if $Z^{(0)}$ arises from the zero-order solution for the macroscopic metric in Eq. (\ref{pertFLRW}), then the differential constraints for $Z$ given in Eq. (\ref{Zdif}) imply that the contribution from the correlation to the zero-order part of the macroscopic field equations must be $O(\epsilon^{1/2})$ or smaller.  Either this, or there must exist as yet unknown solutions of the macroscopic field equations for $Z^{(0)}$ (that is, different to the form of $Z^{(0)}$ given in Section \ref{MGFLRW}, which may be possible but not proven).

\item[2).]  $\Phi=\Psi=0$.  That is, the macroscopic metric in Eq.  (\ref{pertFLRW}) can
only contain spatially homogeneous perturbations (or, at most, the very highly restricted
inhomogeneous perturbations of the form given in Eq.  (\ref{form})).  This would mean that a
macroscopic perturbed FLRW metric with large-scale inhomogeneous terms is not compatible with the
assumptions that have gone into the theory of MG.  In particular, MG may only be applicable
on the largest scales.

\end{itemize}

\subsubsection{Discussion}

 The first of these possibilities would appear to suggest that perturbed FLRW solutions of MG  behave in the same way as
perturbed FLRW solutions of Einstein's equations (to the lowest order of approximation). The MG equations deal with the effects of averaging on macroscopic scales, and this is consequently a statement about the behaviour of space-time at, or above, the averaging scale  (on smaller scales one would expect to have to deal with the linearly perturbed Einstein's equations directly, as usual). This result assumes it is appropriate to apply MG in the way that we have done in this paper, and that new FLRW solutions to the macroscopic field equations that exhibit different behaviour do not remain to be found. If this is the case, then it is suggested that on macroscopic scales the FLRW solutions to Einstein's equations are a valid description of the averaged geometry (possibly with an $O(\epsilon^{1/2})$ renormalization of the spatial curvature). This could be regarded as a structural stability result for the FLRW solutions of Einstein's equations.

The second possibility would appear to restrict the domain of applicability of MG to macroscopic geometries in which the inhomogeneous perturbations are severely restricted, or to situations in which they are spatially homogeneous and isotropic (in the simple way that the theory has been applied here, at least). This still leaves the possibility of phenomena such as  a time-dependent but spatially homogeneous anisotropic shear or tilt when describing the universe on macroscopic scales. Such effects would appear to be consistent with the anisotropies that arise from averaging the flow of matter fields in studies of cosmological backreaction using perturbation theory (to lowest order in the shear anisotropy that characterizes the anisotropy of the flow) \cite{Marozzi}, as well as those that arise from a detailed analysis of the observed large scale anisotropy of the Hubble flow relative to the CMB rest frame \cite{Wiltshire}.

It is clearly on large scales that the effects of averaging are most important, and MG can potentially deal with this if perturbations to the macroscopic FLRW geometry are spatially homogeneous. Again, on scales below the size of the averaging domains one would expect to be able to use the linearly perturbed Einstein's equations directly. What we have found here, however, perhaps suggests that linear perturbation theory is not a valid approach on macroscopic scales when $\mathcal{A}=O(1)$.

\section{Distance Measures in the Macroscopic Universe}
\label{Obs}

Measures of distance are crucial to observational cosmology, as they are often directly
linked to astrophysical observables such as supernovae and the CMB.  Here we will make a first
step towards calculating angular diameter distances and luminosity distances in the FLRW
solutions of MG.

\subsection{The Macroscopic Optical Equations}

As is usual in cosmology, we will make use of the geometric optics approximation.  The behaviour of a bundle of null geodesics in a general space-time is then given by the Sachs optical equations, which read \cite{sachs}
\bea
\frac{d \tilde{\theta}}{d \lambda} +\tilde{\theta}^2-\omega^2 + \sigma^*
\sigma &=& \frac{1}{2} r_{\al \beta} k^{\al} k^{\beta}
\label{sachs1}\\
\label{sachs2}
\frac{d \omega}{d \lambda} + 2 \omega \tilde{\theta} &=& 0\\ \label{sachs3}
\frac{d \sigma}{d \lambda} +2 \sigma \tilde{\theta} &=& -c_{\al \beta \ga \de}
(t^*)^{\al} k^{\beta} (t^*)^{\ga} k^{\de},
\eea
where $\si$ is the complex shear scalar, $\omega$ is the vorticity
scalar, and $\theta$ is the expansion scalar.  The 4-vector $k^a$ is
tangent to the null curves, and the $t^a$ are complex screen vectors that are
null, orthogonal to $k^a$, with unit magnitude, and that are parallel transported along the curves.  The variable $\lambda$ denotes an affine parameter used to measure position along the null curves.

In the usual application of Eqs. (\ref{sachs1})-(\ref{sachs3}) it is common to take $r_{\al \beta}=R_{\al \beta}$ and $c_{\al \beta \ga \de}=C_{\al \beta \ga \de}$, where $R_{\al \beta}$ and $C_{\al \beta \ga \de}$ are the Ricci and Weyl tensors of the microscopic space-time.  Here we are interested in solving Eqs. (\ref{sachs1})-(\ref{sachs3}) using the average values of these quantities.  We will therefore take $r_{\al \beta}=\langle R_{\al \beta} \rangle$ and $c_{\al \beta \ga \de}= \langle C_{\al \beta \ga \de} \rangle$.  By contracting with the 4-vectors $k^{\alpha}$ and $t^{\alpha}$, that have been calculated in the macroscopic geometry, we can then determine estimates for the driving terms in the evolution equations for the expansion and shear of the null congruence (\ref{sachs1}) and (\ref{sachs3}). These, in turn, can be used to infer the behaviour of typical values of $\tilde{\theta}$ and $\sigma$ when making observations over macroscopic scales in cosmology.

Using results from \cite{Zalaletdinov}, we find that
\be
\label{MGricci}
\langle R_{\alpha \beta} \rangle - M_{\alpha \beta} = Q_{\alpha \beta},
\ee
and that
\begin{eqnarray}
\label{MGweyl}
&& \bar{g}^{\al \ta} \left< C_{\al \beta \ga \de} \right> -W^{\ta}_{\ph{\ta}
  \beta \ga \de}\\ \nonumber
&=& Q^{\ta}_{\ph{\ta} \beta \ga \de} - \bar{g}^{\ta}_{\ph{\ta} [ \ga}
  Q^{\si}_{\ph{\si} \de ] \beta \si} + \bar{g}^{\al \ta} \bar{g}_{\beta [ \ga} Q_{\de
  ] \al} +\frac{1}{3} \bar{g}^{\si \rh} Q_{\si \rh} \bar{g}^{\ta}_{\ph{\ta} [\ga}
  \bar{g}_{\de ] \beta}
 + 4 Z^{\ta \ph{\beta [ \ga} \la}_{\ph{\si} \beta [ \ga \ph{\la}
  \uline{\la} \de]}
\\&& \nonumber -Z^{\si \ph{\de [ \beta \ga} \ta}_{\ph{\si} \de [ \beta \uline{\ga}
  \ph{\tau} \si]}
-Z^{\si \ph{ \de [ \beta} \ta}_{\ph{\si} \de [ \beta \ph{\ta} \uline{\ga}
  \si ] }
+Z^{\si \ph{ \al [ \beta \de} \ta}_{\ph{\si} \al [ \beta \uline{\de}
  \ph{\ta} \si]}
+Z^{\si \ph{ \al [ \beta} \ta}_{\ph{\si} \ga [ \beta \ph{\ta} \uline{\de}
  \si]} +Z^{\si}_{\ph{\si} \de [ \al \uline{\ga} \uline{\beta} \si ]} \bar{g}^{\al
  \ta}
+Z^{\si}_{\ph{\si} \de [ \al \uline{\beta} \uline{\ga} \si ]} \bar{g}^{\al
  \ta}
\\&&\nonumber -Z^{\si}_{\ph{\si} \ga [ \al \uline{\de} \uline{\beta} \si ]} \bar{g}^{\al
  \ta}
-Z^{\si}_{\ph{\si} \ga [ \al \uline{\beta} \uline{\de} \si]} \bar{g}^{\al
  \ta} -\frac{1}{3} Z^{\ep \ph{ \si [ \rh} \si \rh}_{\ph{\ep} \si [ \rh
  \ph{ \si \rh} \ep]} \bar{g}^{\ta}_{\ph{\ta} \ga} \bar{g}_{\de \beta}
- \frac{1}{3} Z^{\ep \ph{\si [ \rh} \rh \si}_{\ph{\ep} \si [ \rh \ph{
  \rh \si} \ep]}  \bar{g}^{\ta}_{\ph{\ta} \ga} \bar{g}_{\de \beta}
\\&&\nonumber +\frac{1}{3} Z^{\ep \rh \ph{[ \rh \ga} \ta}_{\ph{\ep \rh} [ \rh
  \uline{\ga} \ph{\ta} \ep]} \bar{g}_{\de \beta}
+\frac{1}{3} Z^{\ep \rh \ph{[ \rh } \ta}_{\ph{\ep \rh} [ \rh
  \ph{\ta} \uline{\ga} \ep]} \bar{g}_{\de \beta} +\frac{1}{3} Z^{\ep \rh}_{\ph{\ep \rh} [ \rh \uline{\beta} \uline{\de}
  \ep]} \bar{g}^{\ta}_{\ph{\ta} \ga}
+\frac{1}{3} Z^{\ep \rh}_{\ph{\ep \rh} [ \rh \uline{\de} \uline{\beta}
  \ep]} \bar{g}^{\ta}_{\ph{\ta} \ga}
+\frac{1}{3} Z^{\ep \ph{\si [ \rh} \si \rh}_{\ph{\ep} \si [ \rh
  \ph{\si \rh} \ep]} \bar{g}^{\ta}_{\ph{\ta} \de} \bar{g}_{\ga \beta}
\\&& \nonumber +\frac{1}{3} Z^{\ep \ph{\si [ \rh} \rh \si}_{\ph{\ep} \si [ \rh
  \ph{\si \rh} \ep]} \bar{g}^{\ta}_{\ph{\ta} \de} \bar{g}_{\ga \beta} -\frac{1}{3} Z^{\ep \rh \ph{[ \rh \de} \ta}_{\ph{\ep \rh} [ \rh
  \uline{\de} \ph{\ta} \ep]} \bar{g}_{\ga \beta}
-\frac{1}{3} Z^{\ep \rh \ph{[ \rh } \ta}_{\ph{\ep \rh} [ \rh
  \ph{\ta} \uline{\de} \ep]} \bar{g}_{\ga \beta}
-\frac{1}{3} Z^{\ep \rh}_{\ph{\ep \rh} [ \rh \uline{\beta} \uline{\ga}
  \ep]} \bar{g}^{\ta}_{\ph{\ta} \de}
-\frac{1}{3} Z^{\ep \rh}_{\ph{\ep \rh} [ \rh \uline{\ga} \uline{\beta}
  \ep]} \bar{g}^{\ta}_{\ph{\ta} \de},
\end{eqnarray}
where $W^{\ta}_{\ph{\ta} \beta \ga \de}$ is the macroscopic Weyl tensor, constructed from $M^{\ta}_{\ph{\ta} \beta \ga \de}$ and $\bar{g}_{\alpha \beta}$ in the usual way. If the macroscopic geometry is FLRW, then we automatically have $W_{\al \beta \ga \de}=0$, so that the average of the Weyl tensor of the microscopic geometry is given only by the RHS of Eq. (\ref{MGweyl}).  The raising and lowering of indices in this equation are done using $\bar{g}^{\al \beta}$ and $\bar{g}_{\al \beta}$.

\subsection{Optics in Type I Solutions}

Without loss of generality, we choose our coordinate system such that the light rays we consider are propagating in the $x$-direction. We also take the tangent vectors in the macroscopic geometry, $k^{\mu}$, to be past directed.  The null and geodesic conditions, $k^{\mu} k_{\mu}=0$ and $k^{\mu} k^{\nu}_{\ph{\nu} ; \mu}=0$, then give
\be
\label{ksol1}
k^{\mu} = \frac{1}{a^2} (-1,1,0,0),
\ee
where we have taken the macroscopic geometry to be a spatially flat FLRW solution of the macroscopic field equations, and we have used the coordinate system $\{\eta,x,y,z\}$, where $\eta$ is conformal time (as used in Section \ref{MGpert}).  A suitable choice of complex screen vectors is then
\be
\label{tsol1}
t^{\mu} = \frac{1}{\sqrt{2} a} ( 0,0,1,-i),
\ee
which are unique up to a rotation in the $y$-$z$ plane.

Let us now use these expressions, together with Eqs. (\ref{MGricci}) and (\ref{MGweyl}), to calculate the driving term on the right-hand side of Eqs. (\ref{sachs1}) and (\ref{sachs3}). These are
\be
\label{rhs1}
-\frac{1}{2} \langle R_{\alpha \beta} \rangle k^{\alpha} k^{\beta} = \frac{1}{a^4} \left[ 2 \frac{a^{\prime 2}}{a^2}- \frac{a^{\prime \prime}}{a} + \mathcal{A}^2 \right]
\ee
and
\be
\label{rhs2}
- \langle C_{\alpha \beta \gamma \delta} \rangle (t^*)^{\alpha} k^{\beta} (t^*)^{\gamma} k^{\delta} = 0,
\ee
where, as before, a prime denotes differentiation with respect to $\eta$.
That is, the shear scalar has no driving term, just as in the FRLW solutions of Einstein's equations, and the total effect of the correlation terms is the presence of $\mathcal{A}/a^4$ on the RHS of Eq. (\ref{sachs1}).

Just as in the macroscopic field equations for spatially homogeneous and isotropic
geometries, we have that $\mathcal{A}$ is the only constant that enters into the final
equations.  What is more, we note that $\mathcal{A}$ once again takes exactly the same
role as a spatial curvature term.  That is, if we were to calculate the right-hand side of
Eqs.  (\ref{sachs1}) and (\ref{sachs3}) in a space-time whose microscopic geometry was given
by spatially curved FLRW, then we would arrive at exactly Eqs.  (\ref{rhs1}) and
(\ref{rhs2}), with spatial curvature $\kappa=\mathcal{A}^2$.  Angular diameter distances and luminosity distances calculated in this way are therefore
identical to those obtained in a spatially curved FLRW solution of Einstein's equations,
even though the macroscopic geometry (as found in \cite{vandenHoogen2009} and \cite{ColeyPelavasZalaletdinvov2005}) is that of a spatially flat FLRW universe.

\subsection{Optics in Type II Solutions}

Let us now consider the RHS of Eqs. (\ref{sachs1}) and (\ref{sachs3}) in the
more general Type II solutions given in Section \ref{TypeII}.  This time we cannot choose light rays that propagate in the $x$-direction without losing generality. We therefore consider rays with tangent vector
\be
k^{\mu} = \frac{1}{a^2} \left( -1 , \cos \theta \cos \psi , \cos \theta \sin \psi, - \sin \theta \right),
\ee
which is simply a spatial rotation of the 4-vector $k^{\mu}$ given in Eq. (\ref{ksol1}).  Here $\theta$ denotes rotation about the $y$-axis, and $\psi$ denotes rotation about the $z$-axis.  Applying this same rotation to the screen vectors from Eq. (\ref{tsol1}) gives
\be
t^{\mu} = \frac{1}{\sqrt{2} a} \left( 0, -\sin \psi -i \cos \psi \sin \theta, \cos\psi -i\sin\psi \sin \theta, -i \cos\psi \cos\theta \right).
\ee
Using these expressions we then find that
\be
\label{rhs1b}
-\frac{1}{2} \langle R_{\alpha \beta} \rangle k^{\alpha} k^{\beta} = \frac{1}{a^4} \left[ 2 \frac{a^{\prime 2}}{a^2}- \frac{a^{\prime \prime}}{a} + \mathcal{A}^2 - \frac{\mathcal{F}_2(x,y)}{2} \cos^2 \theta \cos (2 \psi) \right]
\ee
and
\be
\label{rhs2b}
- \langle C_{\alpha \beta \gamma \delta} \rangle (t^*)^{\alpha} k^{\beta} (t^*)^{\gamma} k^{\delta} = - \frac{\mathcal{F}_2(x,y)}{2 a^4} \cos^2 \theta \left[ \cos (2\psi) +i \sin (2\psi) \sin\theta \right].
\ee
These two equations are identical to those of the Type I solutions, given in Eqs.
(\ref{rhs1}) and (\ref{rhs2}), except for the presence of the terms involving
$\mathcal{F}_2(x,y)$.

\subsubsection{Discussion}

These calculations allow us to consider the effect of inhomogeneities
on geometric optics, and hence cosmological observations, within the framework of MG.
The effects, which can in principle depend on both $x$ and $y$, depend on the direction of the null vector
$k^{\mu}$ and can contribute possible effects both along the null ray and transverse to it. Note that when the null
ray is entirely orthogonal to the $xy-$plane, so that $\theta=\pm \pi/2$,
there is no additional affect.

If either $\theta=0$ or $\psi=0$ then the extra terms involving $\mathcal{F}$ on the
right-hand side of Eqs.  (\ref{rhs1b}) and (\ref{rhs2b}) are identical.  This means that if
expansion is generated from a non-zero $\mathcal{F}$, then the term containing $\mathcal{F}$
in Eq.  (\ref{rhs2b}) will also necessarily drive the shear scalar.  If $\psi=\pi/4$,
however, then the extra term in Eq.  (\ref{rhs1b}) vanishes, while the imaginary part of the
extra term in Eq.  (\ref{rhs2b}) is non-zero.  This means that it is also possible to have a
driving term in the evolution equation from the shear scalar, while not having one in the
evolution equation for the expansion scalar.  In either case, the presence of the terms
involving $\mathcal{F}_2(x,y)$ in these equations breaks the degeneracy between the
contribution from the correlation tensor and the inclusion of a spatial curvature term.

As we can choose $\mathcal{F}_2(x,y)$ to be any function of $x,y$ we like, the amount of freedom in the optical properties of these solutions is significant.   For example, if $\theta=\psi=0$, then for $\mathcal{F}_2(x,y)>0$ the effect of the extra term is to make distant objects appear brighter.  This will be due to both the direct consequences of $\mathcal{F}_2(x,y)$ appearing on the RHS of Eq. (\ref{sachs1}), as well as indirectly, through its contribution to the evolution of the shear in Eq. (\ref{sachs3}).  The former of these acts in the same way as Ricci focussing, while the latter focusses light rays through the appearance of $\sigma^* \sigma$ in Eq. (\ref{sachs1}).

If $\theta=\psi=0$ and $\mathcal{F}_2(x,y)<0$, however, the situation will be somewhat more complicated, with the extra term on the RHS of Eq. (\ref{rhs1b}) causing de-focusing of the light rays, while the contribution from the extra term on the right-hand side of Eq. (\ref{rhs2b}) will still cause $\sigma^* \sigma$ to grow, and hence will cause focusing.  It can then be the case the de-focusing occurs at smaller distances (before the shear has had time to accumulate), while focusing will occur at larger distances (after the shear scalar has had sufficient time to grow large, as is generically expected to happen at large distances in inhomogeneous space-times).  For $\mathcal{F}_2(x,y)<0$ we can therefore have relatively nearby objects appearing dimmer, while more distant objects appear brighter.

The exact optical properties will of course depend on the precise form of the function $\mathcal{F}_2(x,y)$, as well as the values of $\theta$ and $\psi$.  Clearly, if $\mathcal{F}_2(x,y)$ is allowed to have different signs in different regions of the universe, or $\theta$ or $\psi$ are allowed to be non-zero, then the results could be still more complicated.

\section{Concluding Remarks}

In this paper we have discussed the dynamical evolution equations of the Universe on large scales using the theory of Macroscopic Gravity (MG). This theory attempts to model the effects of averaging the geometry of space-time and is consequently important for the interpretation of cosmological observations. In particular, we have investigated spatially homogeneous and isotropic solutions to the MG field equations, and presented new exact FLRW cosmological solutions. In these FLRW cosmological  MG solutions, the macroscopic geometry typically evolves like the FLRW solutions of Einstein's equations but with an extra spatial curvature term in the macroscopic equivalent of the Friedmann equation, even though the macroscopic geometry is spatially flat.

We then investigated perturbations in the macroscopic geometry around these FLRW solutions. We assumed that the scale of these perturbations is larger than that of the averaging domains, that the correlation tensor $Z^{\alpha\phantom{\beta\mu}\gamma}_{\phantom{\alpha}\beta\mu\phantom{\gamma}\delta\nu}$ can be perturbed as in Eq.  (\ref{Zpert}), and that the MG approach is generally valid. We found that the macroscopic metric in Eq.  (\ref{pertFLRW}) can
only contain spatially homogeneous perturbations (or, at most, the very highly restricted inhomogeneous perturbations of the form given in Eq.  (\ref{form})). However, as noted earlier, this still leaves the interesting possiblity of time-dependent but spatially homogeneous effects such as anisotropic shear or tilt. These are permitted on macroscopic scales within MG, and might potentially be of interest due to recent studies of cosmological backreaction such as \cite{Marozzi} and \cite{Wiltshire}.

We then  took a first step towards calculating distance measures using the FLRW solutions of MG we previously obtained. This was done using the geometric optics approximation, and by substituting the averages of the Ricci and Weyl curvature tensors into the driving terms of the Sachs optical scalar equations. We found that in the simplest solutions the correlation tensor acts as a spatial curvature term in the Sachs equations, in the same way that it does in the MG field equations. In more generality, we found that the optical properties depend on terms involving the free function  $\mathcal{F}_2(x,y)$. These terms appears on the right-hand side of Eqs.  (\ref{rhs1b}) and (\ref{rhs2b})), and break the degeneracy between the contribution from the correlation tensor and the inclusion of a spatial curvature term. They also depend on the direction of the null rays to which $k^{\mu}$ is tangent.

Recent studies on the effect of inhomogeneities on the optical properties of universes that are statistically spatially homogeneous and isotropic on large scales have suggested that their consequences can be large enough to be of importance for interpreting observational data (see, for example, \cite{Bolejko}). The motivation for the current work is to start detailed and explicit computations of the effects of inhomogeneities on cosmological observables, such as the growth of large-scale structure and the propagation of light rays, within the framework of MG.

\section*{Appendix A: Type II Macroscopic FLRW Solutions}

Here we shall generalize and expand upon the solution to the MG equations found by van den Hoogen in \cite{vandenHoogen2009}.  To recap, the assumptions made in \cite{vandenHoogen2009} are
\begin{itemize}
\item[1:] We macroscopic metric ${\bar g\,}_{\alpha\beta}$ is the assumed to obey the splitting rule $\langle g_{\alpha\beta}\gamma^{\gamma}{}_{\delta\epsilon} \rangle=\bar{g}_{\alpha\beta}\overline{\gamma}^{\gamma}{}_{\delta\epsilon}$.
\item[2:] The line-element of the macroscopic geometry can be written
$$
ds^2 = \bar g_{\alpha\beta}dx^\alpha dx^\beta = a^2(\eta)[-d\eta^2+dx^2+dy^2+dz^2],
$$
and there exists a time-like unit vector field orthogonal to the spatial hyper-surfaces of homogeneity and isotropy, $u^\alpha = \frac{1}{a(\eta)}(-1,0,0,0)$.
\item[3:] The averaged microscopic energy-momentum tensor can be modeled macroscopically as a perfect fluid with energy density $\rho$ and pressure $p$.
\item[4:] The electric part of the connection correlation tensor is zero, i.e. $Z^{\alpha\ph{\beta\gamma}\mu}_{\ph{\alpha}\beta\gamma \ph{\mu}\nu\sigma }u^{\sigma}=0$,
\end{itemize}
and, finally,
\begin{itemize}
 \item[5:] The connection correlation tensor $Z^{\alpha\ph{\beta\gamma}\mu}_{\ph{\alpha}\beta\gamma \ph{\mu}\nu\sigma }$ and the affine deformation tensor $A^{\alpha}{}_{\ph{\alpha}\beta\gamma}$ are invariant under the same ${\mathcal G}_{6}$ of Killing vectors as the macroscopic metric.
\end{itemize}
We note that if one interprets the contribution of the connection correlations in the MG equations to be an effective fluid with isotropic pressure $p_{grav}$ and energy density $\rho_{grav}$, then assumption 4 implies $\rho_{grav}+3p_{grav}=0$. In what follows we shall assume only that the first 4 assumptions listed above hold true, while relaxing assumption 5.

In general, without taking into account the cyclic constraint, the connection correlation tensor has 720 independent components.  Assumption 4 immediately sets 360 of them to zero. We shall label the remaining 360 variables as $f_i(x,y,z,\eta)$. This assumption is critical to solving the quadratic constraint (\ref{ZZ}). The equi-affine constraint and the cyclic identity equations (\ref{equi-Z}) and (\ref{cyclic}) then yield 239 constraints, leaving 121 independent variables remaining.   The integrability condition (\ref{ZM}) then yields an additional 52 constraints, leaving 69 independent variables remaining.  It is interesting to note at this point that the polarization tensor $Q^\alpha_{\ph{\alpha}\beta\mu\nu}$  contains only 18 of these 69 independent variables, and $Q^\alpha_{\ph{\alpha}\beta\mu\alpha}=Q_{\beta\mu}$ its trace, contains only 6.  The differential equations (\ref{Zdif}) now constrain the remaining independent 69 variables of $Z^{\alpha\ph{\beta\gamma}\mu}_{\ph{\alpha}\beta\gamma \ph{\mu}\nu\sigma }$ to be functions of $x,y$ and $z$ only.  The remaining equations yield a set of 40 linearly independent differential equations in the spatial variables.

We shall now replace assumption 5 with the following set of less restrictive assumptions:
\begin{itemize}
\item[{5a:}] Only $8 \pi G T^{grav}{}^\epsilon_\gamma = -(Z^{\epsilon}_{\ph{\epsilon} \mu \nu\gamma} - \frac{1}{2} \delta^{\epsilon}_{\ph{\epsilon} \gamma}Q_{\mu \nu} ) \bar{g}^{\mu \nu}$, is invariant under the ${\mathcal G}_{6}$ group of motions.
\item[{5b:}] The 40 independent variables $A^\alpha_{\ph{\alpha}\beta\gamma}$ are all assumed to be independent of $z$.  Eq. (\ref{Adif}) then implies that the 18 independent variables that determine $Q^\alpha_{\ph{\alpha}\beta\mu\nu}$ (a subset of the $f_i(x,y,z)$) are also independent of $z$.
\item[{5c:}] We assume that:
\begin{eqnarray*}
&& A^i_{\ph{i}00}=0,\qquad i\in\{1,2,3\}\\
&& A^0_{\ph{0}i0}=0,\qquad i\in\{1,2,3\}\\
&& A^0_{\ph{0}ij}=0,\qquad [i,j]\in\{[1,2],[2,3],[3,1]\}\\
&& A^i_{\ph{i}0j}=0,\qquad [i,j]\in\{[1,2],[2,1],[2,3],[3,2],[3,1],[3,1]\}\\
&& A^i_{\ph{i}33}=0,\qquad i\in\{1,2\}\\
&& A^3_{\ph{3}i3}=0,\qquad i\in\{1,2\}\\
&& A^3_{\ph{3}ii}=0,\qquad i\in\{1,2\}\\
&& A^i_{\ph{i}3i}=0,\qquad i\in\{1,2\}\\
&& A^i_{\ph{i}jk}=0,\qquad  [i,j,k]\in\{[1,2,3],[2,3,1],[3,1,2]\}\\
&& A^i_{\ph{i}0i}=A^j_{\ph{j}0j},\qquad          [i,j]\in\{[1,2],[2,3],[3,1]\}\\
&& A^i_{\ph{i}ij}=A^j_{\ph{j}ii},\qquad          [i,j]\in\{[1,2],[2,1]\}
\end{eqnarray*}
\end{itemize}
Assumption 5a yields 5 algebraic constraints and 3 differential constraints, so that there are now 64 independent variables in $Z^{\alpha\ph{\beta\gamma}\mu}_{\ph{\alpha}\beta\gamma \ph{\mu}\nu\sigma }$. We note that $Q^\alpha_{\ph{\alpha}\beta\mu\nu}$ contains only 13 of these 64 independent variables, and $Q_{\beta\mu}$ contains only 6. The number of linear independent differential equations reduces to 37. One may note that a solution to these 37 differential equations is to let each of the remaining independent variables in $Z^{\alpha\ph{\beta\gamma}\mu}_{\ph{\alpha}\beta\gamma \ph{\mu}\nu\sigma }$ be constant.

Assumption 5c yields $30$ constraints on the $A^\alpha_{\ph{\alpha}\beta\gamma}$, leaving 10 independent variables in $A^\alpha_{\ph{\alpha}\beta\gamma}$. Assuming 5b and 5c together, and solving Eqns. (\ref{Adif}) and (\ref{AR}) simultaneously, reveals an additional 11 constraints.  This leaves 53 independent variables in $Z^{\alpha\ph{\beta\gamma}\mu}_{\ph{\alpha}\beta\gamma \ph{\mu}\nu\sigma }$, of which only 2 are found in $Q^\alpha_{\ph{\alpha}\beta\mu\nu}$ and $Q_{\beta\mu}$.  We label these two ``special'' independent variables as $\mathcal{F}_1(x,y)$ and $\mathcal{F}_2(x,y)$. These assumptions also decrease the number of differential equations determining $Z^{\alpha\ph{\beta\gamma}\mu}_{\ph{\alpha}\beta\gamma \ph{\mu}\nu\sigma }$ by 5. Surprisingly, the two functions $\mathcal{F}_1(x,y)$ and $\mathcal{F}_2(x,y)$ are not restricted or determined at all by what is now a set of 32 differential equations for the other 51 variables $f_i(x,y,z)$ in $Z^{\alpha\ph{\beta\gamma}\mu}_{\ph{\alpha}\beta\gamma \ph{\mu}\nu\sigma }$. One can now solve both Eq. (\ref{Adif}) and Eq. (\ref{AR}) to find four families of solutions parameterized by the ordered pair $(\alpha_1,\alpha_2)=(1/2,0),(1/2,1/2),(1,0), (1,1/2)$.
The values for $A^\alpha_{\ph{\alpha}\beta\gamma}$ in these families are
\begin{eqnarray*}
&&A^x_{\ph{x}xx}=(1-\alpha_2)\mathcal{B}_1(x,y)\\
&&A^y_{\ph{y}xx}=(1-\alpha_1)\mathcal{B}_2(x,y)\\
&&A^x_{\ph{x}xy}=(1-\alpha_1)\mathcal{B}_2(x,y)\\
&&A^y_{\ph{y}xy}=\alpha_2 \mathcal{B}_1(x,y)\\
&&A^x_{\ph{x}yy}=\alpha_2 \mathcal{B}_1(x,y)\\
&&A^y_{\ph{y}yy}=\alpha_1 \mathcal{B}_2(x,y)\\
&&A^z_{\ph{z}zz}=\mathcal{B}_3\\
&&A^0_{\ph{0}ii}=\mathcal{A}\\
&&A^i_{\ph{i}i\eta}=-\mathcal{A}\\
&&A^0_{\ph{0}00}=-\mathcal{A},
\end{eqnarray*}
which depend on two constants, $\mathcal{A}$ and $\mathcal{B}_3$, and two functions, $\mathcal{B}_1(x,y)$ and $\mathcal{B}_2(x,y)$. The functions $\mathcal{B}_1(x,y)$ and $\mathcal{B}_2(x,y)$ must satisfy the following system of differential equations, which are the remaining equations from (\ref{Adif}),
\begin{eqnarray}
\label{A_dif2}
(\alpha_2-\alpha_1)\frac{\partial B_1}{\partial y}=\mathcal{F}_1\\
(\alpha_2-\alpha_1)\frac{\partial B_2}{\partial x}=\mathcal{F}_1\\
\alpha_2\frac{\partial B_1}{\partial x}+(\alpha_1-1)\frac{\partial B_2}{\partial y} =\mathcal{F}_2,
\end{eqnarray}
where the functions $\mathcal{F}_1(x,y)$ and $\mathcal{F}_2(x,y)$ are the two ``special'' functions found in the polarization tensor $Q^\alpha_{\ph{\alpha}\beta\mu\nu}$.  If $\alpha_1\not=\alpha_2$ then these three differential equations can be considered as definitions for $\mathcal{F}_1(x,y)$ and $\mathcal{F}_2(x,y)$, given any two functions $\mathcal{B}_1(x,y)$ and $\mathcal{B}_2(x,y)$, that satisfy $\frac{\partial B_1}{\partial y}=\frac{\partial B_2}{\partial x}$.  If $\alpha_1=\alpha_2$, then $\mathcal{F}_1(x,y) = 0$, and the third equation can be interpreted simply as a definition for $\mathcal{F}_2(x,y)$. For completeness, the values of the polarization tensor $Q^\alpha_{\ph{\alpha}\beta\mu\nu}$ are
\begin{eqnarray*}
&&Q^{x}_{\ph{x} xyx}=Q^{y}_{\ph{y} yxy}=\mathcal{F}_1(x,y)\\
&&Q^{y}_{\ph{y} xyx}=-\mathcal{A}^2 +\mathcal{F}_2(x,y)\\
&&Q^{x}_{\ph{x} yxy}=\mathcal{A}^2 +\mathcal{F}_2(x,y)\\
&&Q^{x}_{\ph{x} zzx}=Q^{y}_{\ph{y} zzy}=Q^{z}_{\ph{z} xxz}=Q^{z}_{\ph{z} yyz}=\mathcal{A}^2.
\end{eqnarray*}
We will not explicitly present the correlation tensor $Z^{\alpha\ph{\beta\gamma}\mu}_{\ph{\alpha}\beta\gamma \ph{\mu}\nu\sigma }$ here, as
it has 1584 non-trivial components in the 54 variables $f_i(x,y,z)$, $\mathcal{F}_1(x,y)$, $\mathcal{F}_2(x,y)$ and $\mathcal{A}$.

It should be noted that these solutions give $$8 \pi G T^{grav}{}^\epsilon_\gamma=\frac{{\mathcal A}^2}{a(\eta)^2} {\rm diag}(3,1,1,1),$$ which is exactly the same as the Type I solutions. The contibution of the correlation tensor to the MG field equations is therefore the same for both the Type I and Type II solutions, despite the latter having considerably less symmetry in $Z^{\alpha\ph{\beta\gamma}\mu}_{\ph{\alpha}\beta\gamma \ph{\mu}\nu\sigma }$ and $A^{\alpha}{}_{\ph{\alpha}\beta\gamma}$. Finally, one should also note that it may be possible to find even more solutions if assumptions 5b and 5c are relaxed.

\section*{Acknowledgements}

TC would like to acknowledge the STFC for support, and the Department of Mathematics and Statistics at Dalhousie University for hospitality while some of this work was carried out. RVDH acknowledges the support of research funds from St. Francis Xavier University.  This work was supported, in part, by NSERC of Canada.

\end{document}